\documentclass[aip,jcp,preprint,floatfix]{revtex4-1}
\usepackage{amsmath,amsfonts,amssymb}
\usepackage{graphicx}
\usepackage{color}

\graphicspath{{./}}

\usepackage{color}
\newcommand{\ca}{Ca$^{2+}$}

\begin{document}
\title{
Comparing allosteric transitions in the domains of calmodulin through coarse-grained simulations}

\author{Prithviraj Nandigrami and John J. Portman}

\affiliation{Department of Physics, Kent State University, Kent, OH
  44242 }

\date{\today}

\begin{widetext}
  
\begin{abstract}
  
\noindent
Calmodulin (CaM) is a ubiquitous \ca-binding protein consisting of two
structurally similar domains with distinct stabilities, binding affinities, and
flexibilities.  We present coarse grained simulations that suggest the mechanism
for the domain's allosteric transitions between the open and closed
conformations depend on subtle differences in the folded state topology of the
two domains.  Throughout a wide temperature range, the simulated transition
mechanism of the N-terminal domain (nCaM) follows a two-state transition
mechanism while domain opening in the C-terminal domain (cCaM) involves
unfolding and refolding of the tertiary structure. The appearance of the
unfolded intermediate occurs at a higher temperature in nCaM than it does in
cCaM
\color{black}  consistent with nCaM's higher thermal stability.
Under approximate physiological conditions, the simulated
unfolded state population of cCaM accounts for
10\% of the population with nearly all of the sampled transitions (approximately
95\%) unfolding and refolding during the conformational change.  
\color{black}
Transient unfolding significantly slows the
domain opening and closing rates of cCaM. This 
potentially influences the mechanism of \ca-binding to each domain.
\end{abstract}

\end{widetext}

\maketitle



\section*{Introduction}

Allostery is central to the precise molecular control necessary for protein
function.  Indirect coupling between distant regions of a protein is often
provided through a conformational transition between a ``closed''
(ligand-free) and ``open'' (ligand-bound) structure upon ligation.
NMR experiments that reveal proteins exist in dynamic equilibrium with multiple
conformers\cite{barbato:92,moy:94,evenas:99,evenas:01,volkman:01,henzler-wildman:07a} suggest that a protein's conformational
dynamics in the absence of a ligand plays an essential role in allosteric
regulation.\cite{ma:99,swain:06,henzler-wildman:07,boehr:09}
The functional dynamics of a folded protein occurs near the bottom of the
funneled energy landscape, a part of the landscape
generally more susceptible
to perturbations than the self-averaged kinetic bottleneck that determines the
mechanism of folding.\cite{bryngelson:95} 
This sensitivity, while important for a protein's ability to dynamically respond
to environmental conditions and interaction with ligands, also makes the
prospect of general organizing principles for allostery
problematic.\cite{zhuravlev:10b} In this paper, we explore the sense in which
the summarizing statement that native state topology determines the folding
mechanism of small single domain proteins\cite{baker:00} carries over to
large-scale conformational transitions.

Due in part to limitations on computational timescales, much theoretical work
modeling largescale conformational transitions in proteins has focused on
simplified, coarse-grained models based on the energy basins defined by the open
and closed conformations.  The Gaussian network and related models describe an
allosteric transition as motion along low frequency normal modes of the closed
state conformational basin.\cite{bahar:97,atilgan:01,tama:01,bahar:05} While 
the dynamics about a single free energy minimum offers a natural
rationale and clear description of the collective motions involved in the
conformational change,\cite{yang:07,bahar:10} a minimal model capable of
capturing the transition mechanism must accommodate the change in dynamics as
protein moves between the two distinct meta-stable free energy basins.  
\color{black} Allosteric transitions have been modeled by several different
methods in which two meta-stable basins are coupled through an interpolation
based on its energy.  For example, minimal energy pathways have been
computed for a potential surface based on the strain energies relative to each
minimum conformation to predict the transition
mechanism.\cite{maragakis:05,chu:07,das:14} 
Structure based simulations 
that couple two conformational basins have also been developed to
understand the mechanism of allosteric
transitions.\cite{best:05,okazaki:06,okazaki:08,chen:07,lu:08,yang:08}
Additionally, transition mechanisms have been described in terms of the evolution of each
residue's local flexibility using a coarse grained variational model.
\cite{tripathi:08,tripathi:09,tripathi:11,tripathi:13a}
Itoh and Sasai present an alternative approach to predict 
allosteric transition mechanisms in which 
contacts from two meta-stable structures are treated on equal footing 
rather than through an interpolated energy function.\cite{itoh:10,itoh:11}
\color{black} 

In this paper, we use coupled structure based simulation of the opening
transition in the domains of calmodulin (CaM) to explore how subtle differences
in the native state topology can lead to qualitative changes in the transition
mechanism.  This work is motivated in part by an intriguing theoretical
prediction\cite{tripathi:09} that the domain opening mechanism of the C-terminal
domain (cCaM) involves local partial unfolding and refolding
while the N-terminal domain (nCaM) remains folded throughout the transition.
\color{black} These distinct transition mechanisms are in harmony with the
Itoh and Sasai's model that predicts cCaM has larger fluctuations than nCaM
during domain opening.\cite{itoh:11} \color{black}
Local unfolding in cCaM is found to relieve regions of high local strain during
the transition\cite{tripathi:11} in agreement with the cracking mechanism of
allosteric transitions discussed by Miyashita et al.\cite{miyashita:03,miyashita:05}

\color{black}
CaM is a ubiquitous \ca-binding protein consisting of two structurally similar globular domains
connected by a flexible linker.
Each domain consists of two helix-loop-helix motifs (the EF-hands)
connected by a flexible linker as shown in Fig.\ref{fig:ncam_ccam_structures}.
Although topologically similar, the two CaM domains have distinct
flexibilities, melting temperatures and thermodynamic \ca-binding 
properties.\cite{tsalkova:85,linse:91,sorensen:98,masino:00}
In the absence of \ca, the C-terminal domain is particularly dynamic\cite{tjandra:95} 
and is less stable than the N-terminal domain in the intact protein and
when separated into isolated domains.\cite{browne:97,sorensen:98,masino:00} 
The C-terminal domain, which has a very low 
denaturation temperature, is reported to be considerably 
unfolded under physiological temperature.\cite{rabl:02}
Furthermore, NMR experiments monitoring the open/closed transition 
of isolated cCaM have
revealed local transient unfolding of helix F during domain
opening.\cite{lundstrom:05}

\color{black}
The simulations presented in this paper suggest that over a wide range of
temperatures, domain opening in cCaM involves global unfolding and refolding, while the
unfolded conformations are much less prominent in nCaM's primarily two-state
domain opening mechanism.  The appearance of an unfolded intermediate at a
sufficiently high temperature is expected and has been reported for similar
simulations of the conformational transition of cCaM\cite{chen:07} and the
homologous protein S100A6,\cite{okazaki:06} as well as other
proteins.\cite{best:05,yang:08} Given the structural similarity of the two
domains, it is harder to anticipate that the unfolded ensemble becomes
locally stable at a significantly higher temperature in nCaM than it does in cCaM.
Both the analytic model and simulations suggest that cCaM is more susceptible to
unfolding during domain opening, despite employing very different
approximations.  
Nevertheless, 
the simulated intermediate is globally unfolded in contrast to the local
unfolding predicted by the analytic model.  In terms of the kinetics, global
unfolding and refolding significantly slows the simulated domain opening rate in
cCaM which potentially can bias the partitioning of \ca-binding kinetics between
induced fit and conformational selection for the two domains.

\section*{Methods}

\begin{figure*}
  \begin{centering}
    \includegraphics[]{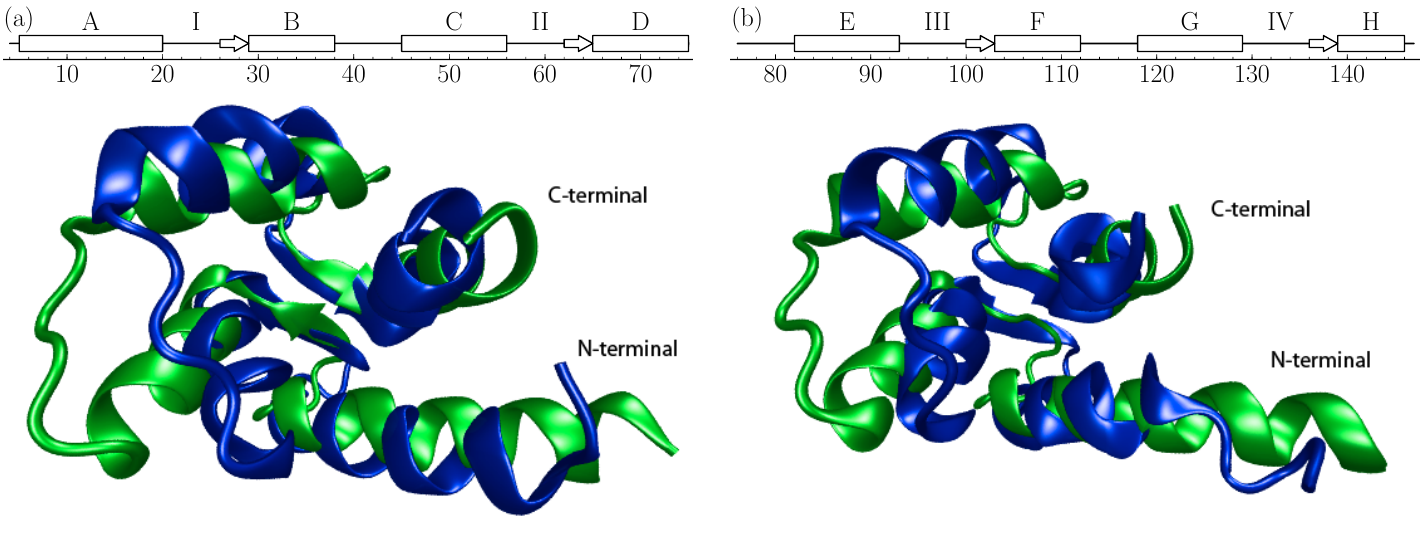}
  \end{centering}
  \caption{ Aligned structures \ca-free (closed/apo) and \ca-bound
    (open/holo) native conformations for (a) N-terminal domain and (b)
    C-terminal domain of Calmodulin.  The closed state (pdb: 1cfd
    \cite{kuboniwa:95} ) is shown in blue, and the open state
    (pdb: 1cll\cite{chattopadhyaya:92}) is shown in
    green.  The closed (apo) and open (holo) conformations of (a) nCaM (residue index 4--75)
    consist of helices A, B and C, D with binding loops I and II
    respectively.  The closed (apo) and open (holo) conformations of
    (b) cCaM (residue index 76--147) consist of helices E, F and G, H with
    binding loops III and IV respectively.  Secondary structure legend
    for nCaM and cCaM are shown on top of the protein structures.
    The CaM structures were made using visual molecular dynamics.\cite{Humphrey:96}}
  \label{fig:ncam_ccam_structures}
\end{figure*}

We use a native-centric model implemented in
the $\textit{Cafemol}$ simulation package\cite{kenzaki:11} 
to study the
open/closed conformational transitions of the isolated N-terminal and C-terminal
domains of CaM.
This model couples two energy basins, one biased to the open (holo) conformation
and the other to the closed (apo) 
reference conformation.\cite{okazaki:06} The open and closed conformations of the 
domains of CaM are shown in Fig.\ref{fig:ncam_ccam_structures}.

A conformation in this coarse-grained model\cite{okazaki:06} 
is specified by
the $N$ position vectors of the C-$\alpha$ atoms of the protein
backbone, $\mathbf{R} = \{ \mathbf{r}_1, \cdots \mathbf{r}_N \}$.  For
an energy basin biased to the reference conformation, $\mathbf{R}_0$,
the energy of a configuration $\mathbf{R}$ can be written as
\begin{equation}\label{eq:single_basin}
  V_{0}({\mathbf R}) = V_{\mathrm{local}}({\mathbf R|\mathbf R_{0}})
  + V_{\mathrm{n}}({\mathbf R|\mathbf R_{0}})
  + V_{\mathrm{nn}}({\mathbf R|\mathbf R_{0}}).
\end{equation}
The first term in Eq.~\ref{eq:single_basin} defines the coarse-grained
backbone
\begin{align}
  \label{eq:local}
  V_{\mathrm{local}}({\mathbf R|\mathbf R_{0}}) & =
  \sum_{\mathrm{bonds}}K_{b} (b_{i} - b_{i}^0)^{2}
  + \sum_{\mathrm{angles}}K_{\theta} (\theta_{i} - \theta_{i}^0)^{2} \nonumber \\
  & +\sum_{\mathrm{dihedrals}} \left[ K_{\phi}[ 1 - \cos (\phi_{i} - \phi_{i}^0) ] \right.\nonumber \\
  &  \mbox{\hspace{4em}} + \left. K_{\phi}^{(3)}[1 - \cos 3(\phi_{i} - \phi_{i}^0)]\right], \nonumber \\
\end{align}
where $b_i$, $\theta_i$, and $\phi_i$ denote bond lengths, bond
angles, and dihedral angles, respectively. The corresponding values in
the native structure are denoted with a superscript: $b_i^0$,
$\theta_i^0$, and $\phi_i^0$. The non-bonded interaction between
neighboring residues in the native structure (native contacts) have
short-ranged attraction
\begin{equation}
  \label{eq:native}
  V_{\mathrm{n}}({\mathbf R|\mathbf R_{0}}) = 
  \sum_{\substack{i<j-3\\\mathrm{native}\\\mathrm{contacts}}}
  \!\!\!\!\!
  \epsilon_{\mathrm{go}} \left[ 5 \left(\frac{r_{ij}^{0}}{r_{ij}}\right)^{12} 
    - 6 \left(\frac{r_{ij}^{0}}{r_{ij}}\right)^{10} \right] ,
\end{equation}
while non-native contacts are destabilized through a repulsive
potential
\begin{equation}
  \label{eq:nonnative}
  V_{\mathrm{nn}}({\mathbf R|\mathbf R_{0}})
  = \sum_{\substack{i<j-3\\
      \mathrm{non-native}}}
  \!\!\!\!
  \epsilon_{\mathrm{rep}} \left(\frac{d}{r_{ij}}\right)^{10}.
\end{equation}
Here, $r_{ij}$ is the distance between C-$\alpha$ atoms $i$ and $j$ in a conformation, $\mathbf{R}$,
and $r_{ij}^0$ is the corresponding separation distance found in the
reference structure, $\mathbf{R}_0$.

The coefficients defining the energy function are set to their default values in
\textit{Cafemol}: $K_{b} = 100.0$, $K_{\theta} = 20.0$, $K_{\phi}^{(1)} = 1.0$
and $K_{\phi}^{(3)} = 0.5$, $\epsilon_{\mathrm{go}}=0.3$,
$\epsilon_\mathrm{rep} = 0.2$ in units of $\mathrm{kcal/mol}$, and $d = 4$\AA.
Trajectories are simulated using Langevin dynamics with a friction coefficient
of $\gamma = 0.25$ and a timestep of $\Delta t = 0.2 $ (in coarse-grained
units).\cite{okazaki:08} With these parameters, the folding transition
temperatures of the isolated CaM domains are estimated from equilibrium
trajectories to be $T_\mathrm{F}^\mathrm{o}(\mathrm{nCaM}) = 333.6^\circ$K and
$T_\mathrm{F}^\mathrm{c}(\mathrm{nCaM}) = 328.9^\circ$K for the open and closed state of nCaM,
and $T_\mathrm{F}^\mathrm{o}(\mathrm{cCaM}) = 335.1^\circ$K and
$T_\mathrm{F}^\mathrm{c}(\mathrm{cCaM}) = 330.5^\circ$K for the open and closed state of cCaM,
respectively.  
\color{black} %
Experimentally, the isolated domains have similar folding transition
temperatures of approximately $323^\circ$K\cite{masino:00}.  Although
$T_\mathrm{F}^\mathrm{o}(\mathrm{cCaM})$ and
$T_\mathrm{F}^\mathrm{o}(\mathrm{nCaM})$ as well as
$T_\mathrm{F}^\mathrm{c}(\mathrm{cCaM})$ and
$T_\mathrm{F}^\mathrm{c}(\mathrm{nCaM})$ are within 2$^\circ$K (with cCaM's
thermal stability slightly below nCaM's ), coupling the open and closed basins
significantly destabilizes cCaM with respect to nCaM (described below).
Consequently, the simulations relevant to the domains of intact CaM, for which
interactions between the domains, particularly with the linker
region\cite{odonnell:09}, reduce the folding temperature of the C-terminal
domain to roughly 315$^\circ$K and increase the folding temperature of
N-terminal domain to 328$^\circ$K.\cite{tsalkova:85,sorensen:98,masino:00}
\color{black}

To study conformational changes between two meta-stable states, the energies of
the corresponding native basins, $V_1(\mathbf{R})$ and $V_2(\mathbf{R})$, are
coupled through an interpolation function\cite{kenzaki:11}
\begin{equation}\label{eq:coupled_go_model_eqn}
  V({\mathbf R}) = \frac {V_{1}+V_{2}+\Delta V}{2}-\sqrt {\left(\frac {V_{1}-V_{2}-\Delta V}{2}\right)^2 + \Delta ^2}.
\end{equation}
Here, the interpolation parameters, $\Delta$ and $\Delta V$, control the barrier
height and the relative stability of the two basins.  The single basin energies
$V_1(\mathbf{R})$ and $V_2(\mathbf{R})$ are computed from
Eq.~\ref{eq:single_basin} with modifications to some of the reference parameters
in the potential in order to minimize conflicts between the two contact maps.
(See Ref.~\onlinecite{okazaki:06,okazaki:08,kenzaki:11} for details).

\begin{figure}
  \begin{center}
    \includegraphics[width=2.75in]{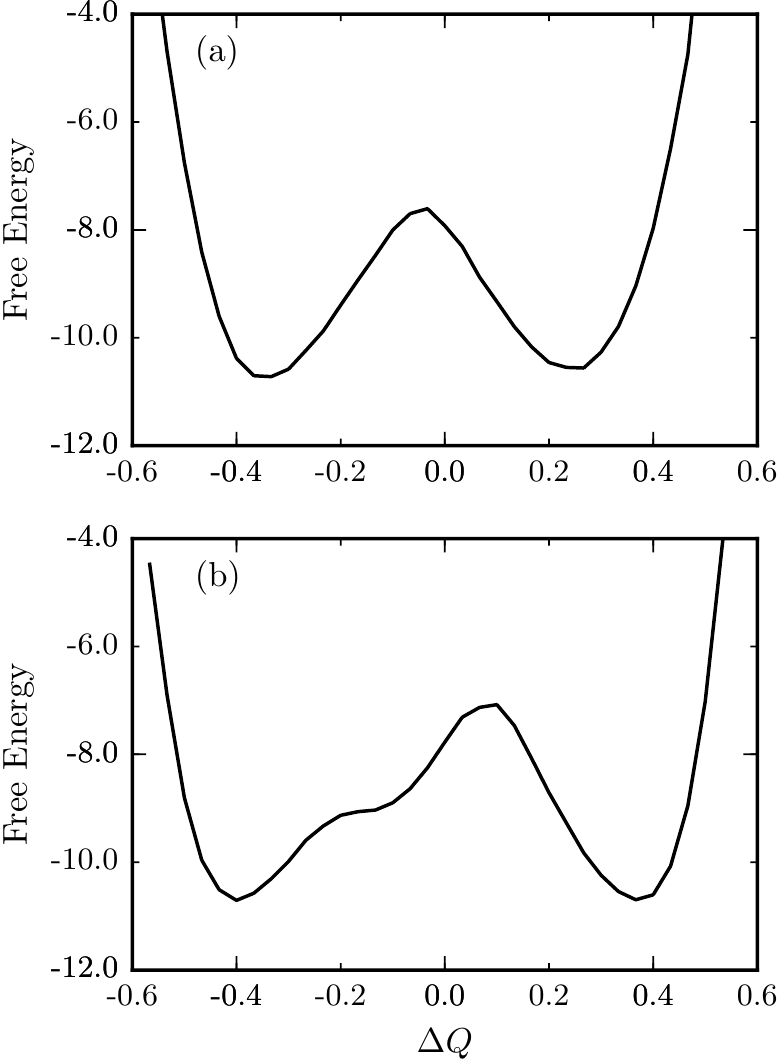}
    \caption{Simulated free energy (in units of $\mathrm{k_{B}T}$) 
      as a function of the global progress coordinate
      $\Delta Q$ = $Q_\mathrm{closed} - Q_\mathrm{open}$
      for (a) nCaM and (b) cCaM. }
    \label{fig:1d_free_energy}
  \end{center}
\end{figure}

\begin{figure}
  \begin{center}
    \includegraphics[width=2.75in]{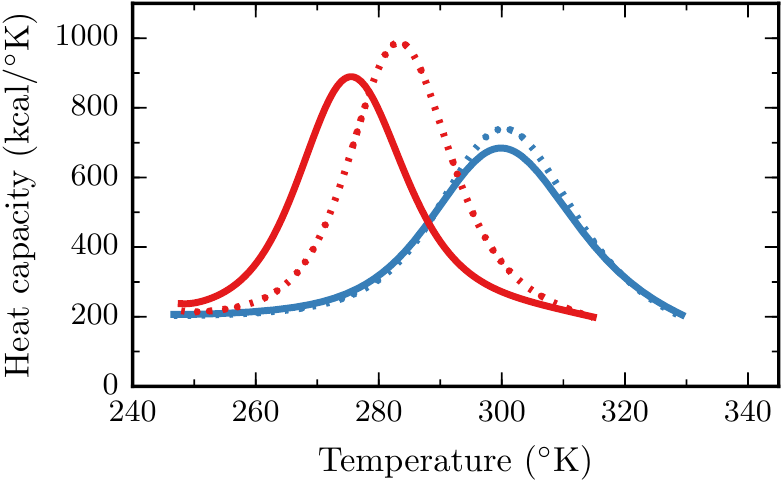}
    \caption{
      \color{black}
      Heat capacity as a function of temperature
      for cCaM (red) and nCaM (blue) for two relative stabilities of the open and closed
      basins. The solid curves correspond to equally stable open and closed basins, and in
      the dashed curves the open state occupies approximately 10\% of the total population. 
      \color{black}
    }
    \label{fig:cv}
  \end{center}
\end{figure}

To compare the simulated domain opening mechanisms  most clearly, it is
convenient to choose coupling parameters $\Delta$ and $\Delta V$ so that the
barrier between the two states is low enough to give sufficient sampling of the
two states and equal stability of the open and closed conformations (a choice to
improve sampling of the equilibrium transition kinetics).  With
$\Delta = 14.0 \,\mathrm{kcal/mol}$ and $\Delta V = 2.15 \,\mathrm{kcal/mol}$
for nCaM, and $\Delta = 17.5 \,\mathrm{kcal/mol}$ and
$\Delta V = 0.25 \,\mathrm{kcal/mol}$ for cCaM, the open and closed states are
equally probable with a free energy barrier of $\simeq 4\mathrm{k_{B}T}$ as
shown in Fig.\ref{fig:1d_free_energy}.
\color{black}
With these parameters, the folding temperature for cCaM is approximately
25 degrees below the folding temperature of nCaM as
indicated by the peaks in the heat capacity shown in Fig.\ref{fig:cv}.  
We report temperatures relative to the simulated folding temperature of cCaM, denoted as
$T_\mathrm{{F}}^{\star}=275.0^\circ$K.  Although we have explored a wide range of
temperatures, most of the results presented in this paper have
$T_\mathrm{{sim}} = 0.96T_\mathrm{{F}}^{\star}$, a temperature slightly below
the folding temperature of cCaM, and significantly below the folding temperature
of nCaM.
\color{black} %

\color{black} 
NMR experiments indicate that the closed state of cCaM is 
more stable than the open state under physiological conditions,
accounting for roughly 90\% of the population.\cite{malmendal:99} Assuming nCaM is
similar, we adjust the relative stability of both domains through the coupling
parameter $\Delta V$ to match this stability
($\Delta V = 3.5 \,\mathrm{kcal/mol}$ for nCaM, and
$\Delta V = 4.0 \,\mathrm{kcal/mol}$ for cCaM).  As shown in Fig.\ref{fig:cv},
the folding temperatures of the domains are sensitive to this destabilization of the
open state.  The simulated folding temperatures of the two domains differ
by approximately $18^\circ$K, somewhat larger than the difference in
experimental folding temperatures of the domains in intact CaM, approximately
$13^\circ$K.\cite{rabl:02} To connect to the domain opening kinetics in intact CaM,
we relate the simulated temperatures to the folding temperatures of its
N-terminal and C-terminal domains. With this choice, the physiological
temperature $310^\circ$K corresponds to simulation temperature of 95\% of nCaM's
folding temperature, and 98\% of cCaM's folding temperature.

\color{black}

Simulated conformational ensembles are characterized through local and
global structural order parameters based on the contacts formed in
each sampled conformation.  
A native contact is considered to be formed 
if the distance between the residues is closer than 1.2 times the corresponding
distance in the native conformation.
To characterize structural changes during the conformational transition, 
it is convenient to separate
the set of native contacts in the open (holo) and closed
(apo) conformations into three groups: those that occur
exclusively in either the open or the closed native reference conformation, 
and those that are common to both states.  For each of these groups,
denoted by $\alpha = $ (open, closed, and $\cap$), we define a local
order parameter, $q_{\alpha}(i)$, as the fraction of native contacts
formed involving the $i^{\mathrm{th}}$ residue. Overall native
similarity is monitored by corresponding global order parameters,
$Q_{\alpha} = \langle q_\alpha(i) \rangle$, where the average is taken
over the residues of the protein. 
The free energy parameterized by these
global order parameters
are used to identify locally stable conformational ensembles such as
the open and closed basins.

The transition rates between two coarse-grained ensembles
are calculated from equilibrium simulations of length $10^8$ steps
typically involving $O(10^3)$ open/closed transitions for nCaM
and $O(10^2)$ open/closed transitions for cCaM.
The transition rate between two states labeled by $i$ and $j$
is estimated by\cite{buchete:08}
\begin{equation}\label{eq:transition_rate0}
k_{i \rightarrow j} 
= \frac{N_{i\rightarrow j}}{\langle \tau_i\rangle \sum_{k\ne i} N_{i \rightarrow k}} ,
\end{equation}
where $\langle \tau_i\rangle$ is the mean time spent in state $i$ between
transitions, and $N_{i\rightarrow j}$ are the number of transitions from state
$i$ to state $j$.  When the allosteric transition involves only the open and
closed states, Eq.~\ref{eq:transition_rate0} reduces to the two state rates,
$k_{o\rightarrow c} = \langle \tau_o\rangle^{-1}$ and
$k_{c\rightarrow o} = \langle \tau_c\rangle^{-1}$, where
$\langle \tau_o \rangle$ and $\langle \tau_c \rangle$ are the mean first passage
times to leave the open and closed state, respectively.

\section*{Conformational transitions of isolated
  domains}

\begin{figure}
  \begin{center}
    \includegraphics[]{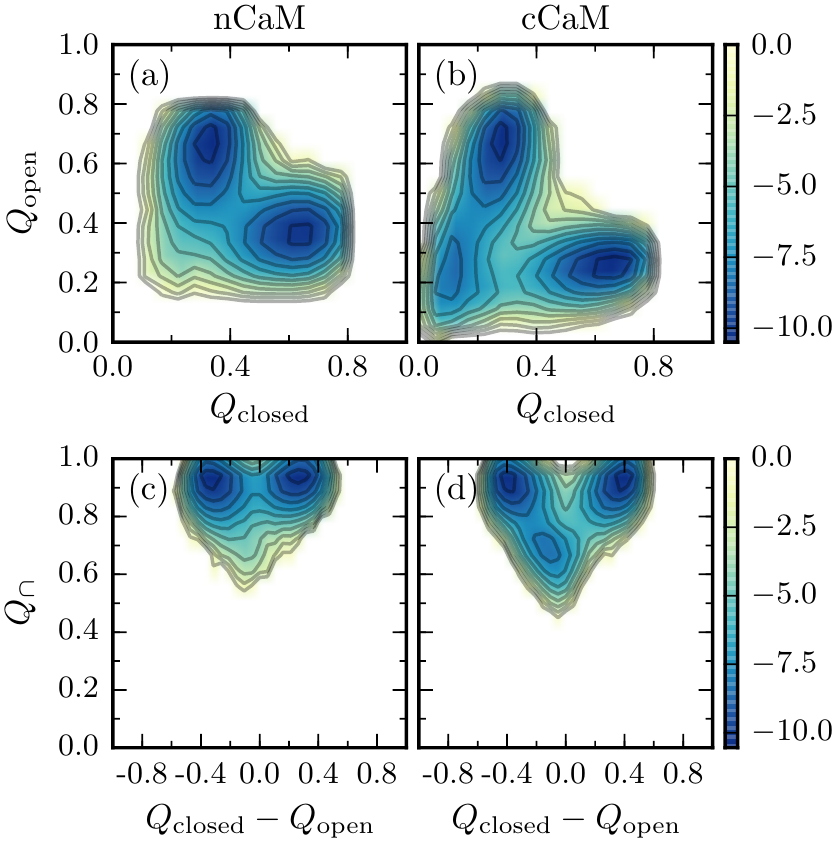}
    \caption{Free energy, in units of $k_\mathrm{B}T$, projected onto
      global order parameters $Q_\mathrm{open}$, $Q_\mathrm{closed}$,
      and $Q_\cap$ for nCaM (a and c) and cCaM (b and d).  The
      intermediate in free energy surface of cCaM corresponds to an
      ensemble of states with intact secondary structure but lacking
      stable tertiary contacts.  }
    \label{fig:Q_plots}
  \end{center}
\end{figure}

The populations of simulated conformations
organized in terms of global order parameters
are shown in Fig.\ref{fig:Q_plots}. The free energy
as a function of $Q_{\mathrm{open}}$ and $Q_{\mathrm{closed}}$  
shows that the nCaM has a two-state domain opening and its conformational
transition is sequential.  That is, contacts specific to the closed conformation
are lost prior to formation of contacts specific to the open conformation which
mostly form after transition state region.  Fig.\ref{fig:Q_plots}
also shows the free energy projected onto the order parameter monitoring common
contacts, $Q_\cap$, and a progress coordinate for the conformational transition,
$\Delta Q = Q_{\mathrm{closed}} - Q_\mathrm{open}$.  The global order parameter
$Q_\cap$ monitors the overall structural integrity of the secondary structure as
well as tertiary contacts within parts of the protein that do not have large
conformational changes during the transition.  As shown in 
Fig.\ref{fig:Q_plots}, the common contacts in nCaM's transition state
ensemble remain largely intact.  In contrast, the simulated open/closed free energy for
cCaM has a locally stable intermediate state.  Simultaneously low values of
$Q_{\mathrm{open}}$ and $Q_{\mathrm{closed}}$ (both less than 0.3), and $Q_\cap$
(less than 0.7), indicate that the intermediate has significantly reduced
tertiary structure.
The probability to form individual contacts in the intermediate (data not shown)
verifies that the secondary structure remains intact, though nearly all the
tertiary interactions are lost.  
Since the barrier for the transition
$\mathrm{closed}\rightarrow \mathrm{I}$
($\Delta F^{\dagger} \simeq 4 \, \mathrm{k_{B}T}$) is higher than the barrier
for $\mathrm{I}\rightarrow \mathrm{open}$
($\Delta F^{\dagger} \simeq 1.5 \, \mathrm{k_{B}T}$), the intermediate can be
considered to be part of cCaM's extended open basin.

\begin{figure}
  \begin{center}
    \includegraphics[]{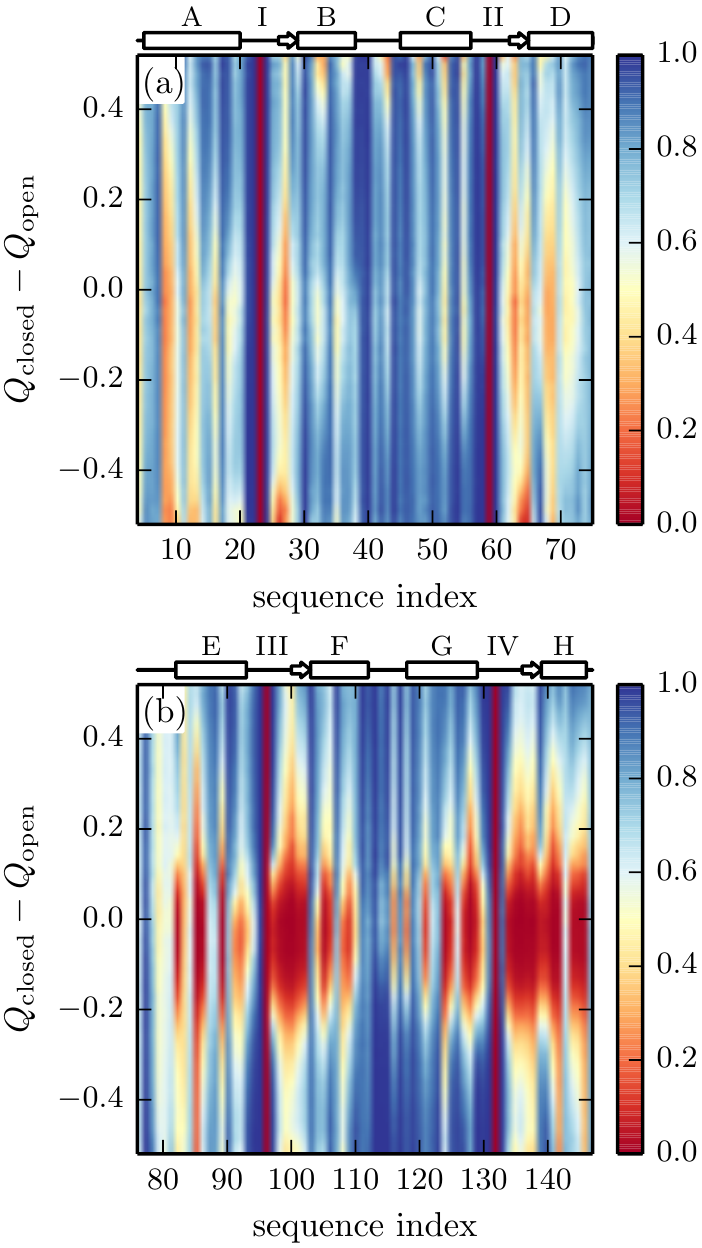}
    \caption{
      Local order parameter, $q_\cap(i)$, plotted as a function of the 
      global progress coordinate, $Q_{\mathrm{closed}} - Q_{\mathrm{open}}$, for each
      residue of (a) nCaM and (b) cCaM. The color represents the
      probability of each residue forming native contacts common to
      both the open and closed structures: low probability is shown
      by red and high probability is shown by blue.  
    }
    \label{fig:q_common}
  \end{center}
\end{figure}

To describe the transition mechanism at the residue level, we consider the local
order parameter $q_\cap(i)$ of each residue as a function of the global progress
coordinate $\Delta Q$.
As shown in Fig.\ref{fig:q_common}, cCaM's residues lose the majority of their
common contacts upon opening (moving upward in the plot) and regain them later
in the transition.  Although the folding and refolding of residues in helices E
and H are more gradual than other residues, nearly every residue (except the
residues in the linker region between helices F and G) looses native tertiary
structure.  In contrast, the common contacts in nCaM remain intact throughout
the transition, though the contacts involving specific residues in helices A and
D and the $\beta$-sheets in the loops are strained. Limited loss of long range
common contacts in nCaM reflect an increased flexibility of the folded transition state
ensemble.

\begin{figure}
  \begin{center}
    \includegraphics[width=2.75in]{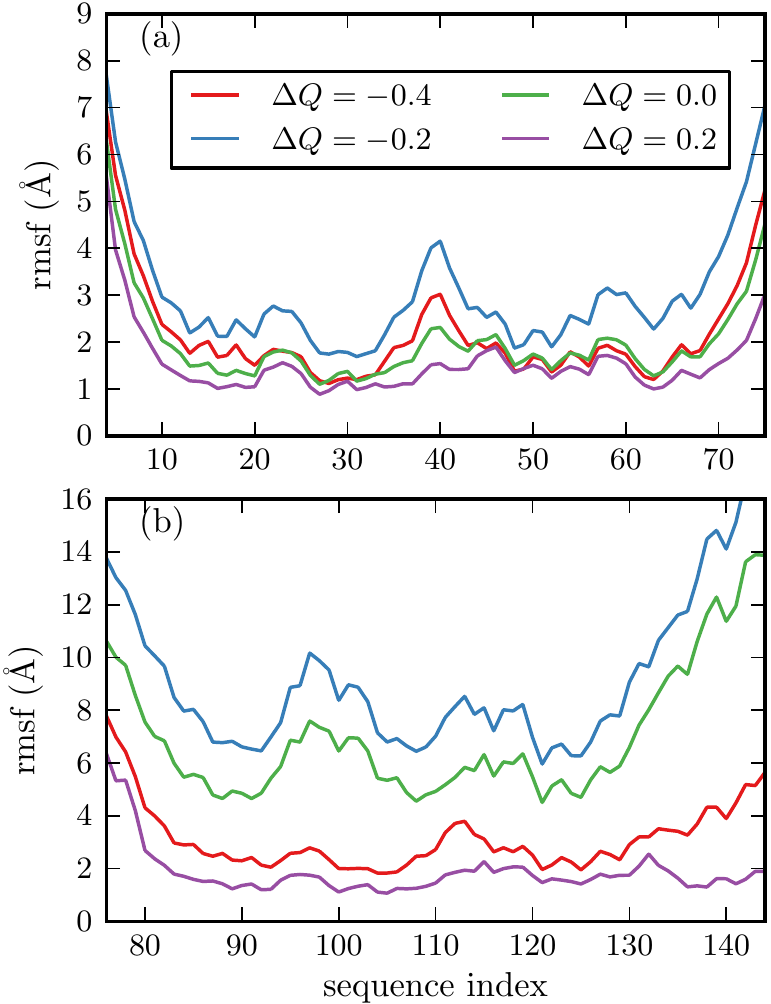}
    \caption{ Magnitude of the root mean square fluctuations for
      each residue for the conformational ensembles along the
      transition pathway for (a) nCaM and (b) cCaM. Each color
      corresponds to the value of $\Delta Q = Q_{\mathrm{closed}} -
      Q_{\mathrm{open}}$ indicated in the legend in (a). }
    \label{fig:rmsf}
  \end{center}
\end{figure}

A coarse-grained, analytic model, 
also predicts distinct transition mechanisms for each domain in which
cCaM is susceptible to local unfolding during the
open/closed transition, while nCaM remains folded.\cite{tripathi:09,tripathi:11}
The conformational transition in the analytic model is described as the evolution 
of local flexibility along the transition route. 
Fig.\ref{fig:rmsf} shows the simulated local flexibility for four discrete values of the
progress coordinate, $\Delta Q$.  Although the fluctuations of the residues in
both domains increase and then decrease during the transition, 
the magnitude of the largest fluctuations are much greater in cCaM.  
In contrast to the global unfolding observed in the
simulations, unfolding and refolding of cCaM predicted by 
the analytic model is localized to particular
residues (primarily in the linker between helix F and G).

\begin{figure}[h]
  \begin{center}
    \includegraphics[width=8.5cm]{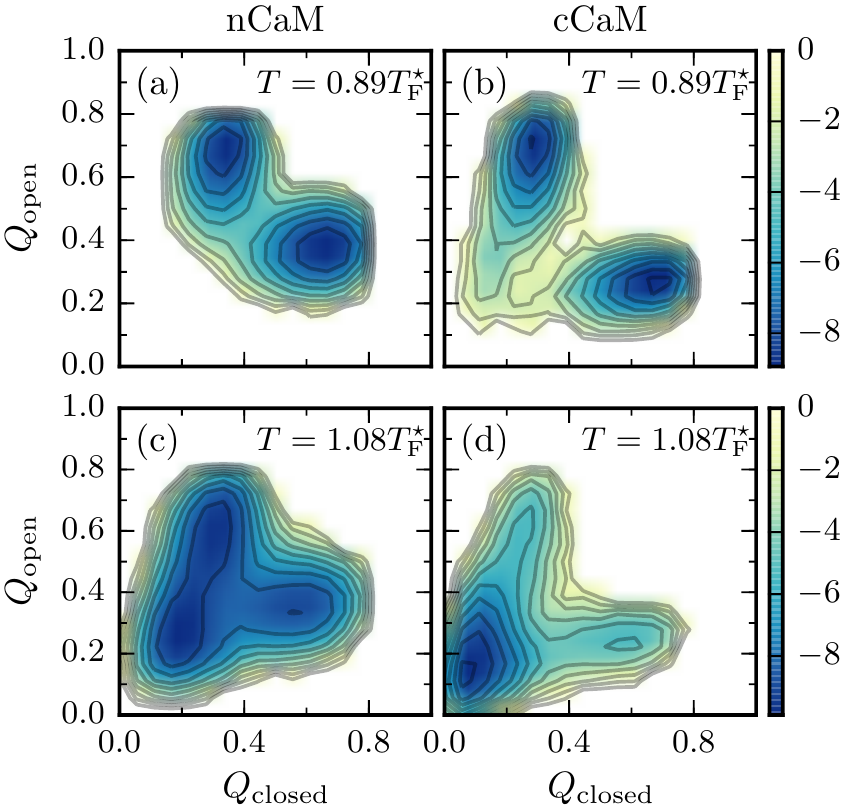}
    \caption{ Free energy, in units of $k_\mathrm{B}T$, projected onto global
      order parameters $Q_\mathrm{closed}$ and $Q_\mathrm{open}$ for nCaM (a and
      c) and cCaM (b and d) with temperature \color{black} %
      $T_\mathrm{{sim}} = 0.89 T_\mathrm{{F}}^{\star}$ and
      $T_\mathrm{{sim}} = 1.08 T_\mathrm{{F}}^{\star}$.  \color{black} %
      At lower temperatures, the unfolded conformations are destabilized so that
      the transition mechanism in both domains becomes more two-state.  At
      higher temperatures, the unfolded states are stabilized for both nCaM and
      cCaM. }
    \label{fig:F_vs_Q_at_different_T}
  \end{center}
\end{figure}

Exploring a range of temperatures reveals that both domains can exhibit a
two-state transition mechanism or a transition mechanism that involves unfolding
and refolding depending on the temperature (see
Fig.\ref{fig:F_vs_Q_at_different_T}).  The transition mechanism at low
temperatures is two state, involving primarily well folded conformational
ensembles throughout the transition. Increasing the temperature progressively
stabilizes the unfolded ensemble until it becomes locally stable at a spinodal
temperature, $T_\mathrm{s}$. Above the spinodal temperature, the transition
between the open and closed state involves unfolding and refolding of the
domain. At high enough temperatures, the unfolded conformation becomes the most
stable state.

Although both domains follow similar transition scenarios as a function of
temperature, the domains can have different transition mechanisms from each
other because the spinodal temperatures are different. Comparing the two
domains, cCaM has a lower spinodal temperature
\color{black}
($T_\mathrm{s}^\mathrm{c} \approx 0.93 T_{\mathrm{F}}^{\star}$) than nCaM
($T_\mathrm{s}^\mathrm{n} \approx 1.005 T_\mathrm{{F}}^{\star}$).  
\color{black} 
For low temperatures, ($T < T_\mathrm{s}^\mathrm{c}$), both the domains have two state
transitions.  For intermediate temperatures
($T_\mathrm{s}^\mathrm{c} < T < T_\mathrm{s}^\mathrm{n}$), the domain opening transition
of nCaM is two state, while the transition of cCaM involves unfolding and
refolding. For higher temperatures ($T_\mathrm{s}^{\mathrm{n}} < T$), the unfolded
ensemble of nCaM is locally stable, but at this temperature the unfolded
ensemble of cCaM is stabilized enough to become the global minimum.

\color{black}
Focusing on the scenario when the open state is 10\% of the total
population and at a simulation temperature corresponding to $T = 310^\circ$K
(to model intact CaM at physiological conditions),
the simulated unfolded population is less than 1\% for nCaM, and approximately
9\% for cCaM. These equilibrium unfolded populations
can be compared to reports of 2\% for the N-terminal
domain and 24\% for the C-terminal domain in intact CaM
based on thermodynamic stability measurements.\cite{masino:00}
\color{black}

\section*{Transition Kinetics}

Using Eq.~\ref{eq:transition_rate0} to calculate opening rates for each domain
at $T_\mathrm{sim}$, we find that unfolding and refolding along the transition
route significantly slows cCaM's domain opening rate compared to
nCaM. Quantitatively, the domain opening and closing rates of nCaM,
$k_{o \rightarrow c} = k_{c \rightarrow o} = 2 \times 10^{-3} \Delta t^{-1}$,
are 50 times larger than the effective opening and closing rates of cCaM,
$k_{o \rightarrow c} = k_{c \rightarrow o} = 4 \times 10^{-5} \Delta t^{-1}$.

A closer look at cCaM's kinetic transitions reveals that only $\approx$ 5\% of
its transition paths proceed through direct transitions from the closed to open
state without significant unfolding along the way.  The rest of the transitions
occur according to the kinetic equation
\begin{equation}\label{eq:ccam_rates}
  \mathrm{closed (c)} \overset{\mathrm{slow}}{\rightleftarrows}
  \mathrm{I} \overset{\mathrm{fast}}{\rightleftarrows}
  \mathrm{open (o) },
\end{equation}
where  $k_{c\rightarrow I} = 4 \times 10^{-5}\Delta t^{-1}$,
$k_{I\rightarrow c} = 2 \times 10^{-4}\Delta t^{-1}$, $k_{I\rightarrow o} = 8
\times 10^{-3}\Delta t^{-1}$, and $k_{o\rightarrow I} = 2 \times 10^{-3}\Delta t^{-1}$
are the corresponding simulated rates between the open, closed, and intermediate states. 

Equilibrium between the open and the unfolded intermediate is established
quickly on the timescale of the conformational transition so that the unfolded
intermediate establishes a steady-state population
\begin{equation}\label{eq:steady_state}
  P_\mathrm{I} = 
  \frac{k_{c\rightarrow I}P_\mathrm{c} + k_{o\rightarrow I}P_\mathrm{o}}
  {k_{I\rightarrow c} + k_{I\rightarrow o}},
\end{equation}
where $P_{c}$ and $P_{o}$ are the equilibrium populations of the closed and open
state respectively.  The effective two-state kinetics for open/closed transition
can be written as
\begin{equation}
  \label{eq:kc_eff}
  k_{c\rightarrow o}^{\mathrm{eff}}
  = \frac{k_{c\rightarrow I}k_{I\rightarrow o}}{k_{I\rightarrow c} 
    + k_{I\rightarrow o}} \\
\end{equation}
and
\begin{equation}
  \label{eq:ko_eff}
  k_{o\rightarrow c}^{\mathrm{eff}}  
  = \frac{k_{o\rightarrow I}k_{I\rightarrow c}}{k_{I\rightarrow c} 
    + k_{I\rightarrow o}} .
\end{equation}

Since $k_{I\rightarrow c} \ll k_{I\rightarrow o}$, these expressions for the
two-state rates can be simplified.  The effective domain opening rate is
determined by the unfolding of the closed state
\begin{equation}\label{eq:effective_rates_c_to_o}
  k_{c \rightarrow o}^{\mathrm{eff}}  \approx k_{c \rightarrow I},
\end{equation}
and the closing rate can be understood through the equilibration of the
intermediate and open state
\begin{equation}
  \label{eq:effective_rates_o_to_c}
  k_{o \rightarrow c}^{\mathrm{eff}} 
  \approx k_{I \rightarrow c}\left( \frac{P_{I}}{P_{o}} \right),
\end{equation}
where $P_{I}/P_{o} = 0.2$ is the population of the unfolded intermediate
relative to the open state.  The simulated effective two state rates for cCaM
are consistent with this steady-state description of the kinetics.

\color{black}
The slowing influence of the folding and unfolding transition persists
when the open state is destabilized to 10\% of the total population, with
domain opening approximately 45 times faster in nCaM than in cCaM
at simulated temperatures that correspond to $T = 310^\circ$K.

\color{black}
\section*{Discussion}

Although the isolated domains of CaM are topologically similar, the simulated
open/closed transition mechanisms are distinct due to the presence of an
unfolded intermediate that appears in the free energy landscape at a different
temperature for each domain. Two-state transition kinetics persist at higher
temperatures in nCaM, whereas the unfolded ensemble is more readily stabilized
in cCaM.  Above the spinodal temperature, transient unfolding and refolding of
the domain occurs through the locally stable unfolded intermediate (exemplified
by cCaM at $T_\mathrm{sim}$).  Below the spinodal temperature, the transition is
two-state like albeit with conformational dynamics that anticipates the unfolded
intermediate with high flexibility and stressed tertiary interactions (as in
nCaM at $T_\mathrm{sim}$).

The unfolding and refolding along the open and closed transition is reminiscent
of the cracking mechanism\cite{miyashita:03,miyashita:05,whitford:07} in which
regions of high local strain are relieved through unfolding and refolding in the
transition region. Since the unfolded conformations involved in cracking are
typically locally unstable, the domain opening of CaM most closely follows this
canonical description at temperatures near the spinodal for the unfolded
conformations.

High temperature unfolded intermediates have been reported previously in
simulations of the open/closed transition in cCaM\cite{chen:07} and the
homologous protein S100A6.\cite{okazaki:06}  
 Chen and Hummer found that the population of the open
ensemble is comparable to that of a marginally stable unfolded ensemble
within a narrow temperature range.  They argue that the sensitive balance
between unstable folding and unfolded populations explains why some experiments
report an open/closed
transition,\cite{evenas:99,malmendal:99,vigil:01,lundstrom:04,lundstrom:05} and
others report folding/unfolding transition for cCaM under similar
conditions.\cite{rabl:02}

Our simulations suggest that subtle differences in the topology and stability of
the two domains can result in distinct transition mechanisms.  In particular, we
find that the unfolded population is stabilized more readily in cCaM, a result
consistent with the prediction that cCaM (and not nCaM) exhibits local folding
and unfolding during opening.\cite{tripathi:08,tripathi:11} The C-terminal
domain's lower spinodal temperature may reflect its decreased overall relative
thermodynamic stability.  Indeed, nCaM is measured to be more stable than cCaM
in the absence of \ca,\cite{sorensen:98} with cCaM being significantly unfolded
at room temperature (20 -- 25$^\circ$C).\cite{masino:00}

The transient unfolding and refolding observed in the simulations significantly
slows the transition kinetics of cCaM.  Several key observations of CaM dynamics
have been reported, but how the dynamics of the individual domains compare
is not clear from the literature.  NMR studies of intact CaM in the absence of
\ca report that cCaM is more dynamic than the nCaM, with an exchange time of $350~\mu\mathrm{s}$ for
cCaM.\cite{tjandra:95}  This timescale is comparable to the
folding and unfolding equilibration time of 200 $\mu\mathrm{s}$ for cCaM under
similar conditions.\cite{rabl:02} The dynamics of \ca-loaded cCaM with a mutation
E140Q that stabilizes the open state and prevents binding to loop IV exhibits
exchange on the faster timescale of $25~\mu\mathrm{s}$\cite{evenas:01} and
undergoes local transient unfolding.\cite{lundstrom:05}  The dynamics of both
domains under similar conditions has been reported by Price and co-workers who
used fluorescence correlation spectroscopy coupled to F{\"o}rster Resonance
Energy Transfer (FRET) to monitor the intramolecular dynamics of both nCaM and
cCaM on the microsecond timescale.\cite{price:11} They report that both domains
have fluctuations on the $30$ -- $40~\mu\mathrm{s}$ timescale in the absence of
\ca. The \ca-dependence of the fluctuation amplitude, however, indicates
that the observed fluctuations couple to the occupancy of the binding sites (and
hence to domain opening) only in nCaM.  Taken together, 
the evidence that the two domains have a different conformational timescale 
and/or mechanism is
intriguing in light of the predictions from the coarse-grained simulations.
Nevertheless, understanding how flexibility and transient unfolding influences
domain opening dynamics of CaM requires further experimental clarification.

\section*{Concluding Remarks}
 
Understanding the open/closed conformational dynamics of CaM is an essential
step towards modeling \ca-binding.  Exploring how transient unfolding in domain
opening of CaM influences ligand binding is particularly interesting.
Simulations of an extension of this model that includes \ca-binding (reported in
a separate publication) shows that the two domains differ significantly in their
thermodynamic properties such as binding affinity and cooperativity.
Nevertheless, these thermodynamic differences seem to depend on the distinct
conformational properties of the open and closed ensembles of each domain rather
than the presence of an unfolded intermediate.  Transient unfolding may still
influence binding kinetics due to the slowing of the domain opening rate.
This is particularly interesting because the detailed binding mechanism, such as
the partitioning of binding kinetics into conformational selected or induced fit binding
routes\cite{hammes:09} is thought to be sensitive to the timescale of the open
and closed transition.\cite{cai:11} Clarifying how the speed of conformational
dynamics influences the kinetic binding mechanism through a molecular model is a
rich problem that we wish to explore in the future.

\begin{acknowledgments}
  We would like to thank Swarnendu Tripathi for interesting discussions, and
  Daniel Gavazzi for help in preparing some of the figures.  Financial support
  from the National Science Foundation Grant No. MCB-0951039 is gratefully
  acknowledged.
\end{acknowledgments}

%

\end{document}